\documentclass[aps,prl,longbibliography,%preprint
,superscriptaddress, twocolumn,notitlepage]{revtex4-1}

\usepackage{lipsum}
\usepackage{amsmath,amssymb}
\usepackage{natbib}
\usepackage{graphicx,subfigure}
\usepackage{hyperref}
\usepackage{mathtools}
\usepackage[normalem]{ulem}
\newcommand{\be}{\begin{equation}}
\newcommand{\ee}{\end{equation}}
\newcommand{\bse}{\begin{subequations}}
\newcommand{\ese}{\end{subequations}}
\newcommand{\ba}{\begin{eqnarray}}
\newcommand{\ea}{\end{eqnarray}}
\newcommand{\bea}{\begin{eqnarray}}
\newcommand{\eea}{\end{eqnarray}}

\usepackage{color}
\usepackage[normalem]{ulem}  

\begin{document}

%\preprint{}

\title{Erasure tolerant quantum memory and the quantum null energy condition in holographic systems}

\author{Avik Banerjee}
\email{avikphys02@gmail.com}
\affiliation{Center for Quantum Information Theory of Matter and Spacetime, and Center for Strings, Gravitation and Cosmology, Department of Physics, Indian Institute of Technology Madras, Chennai 600036, India}
\author{Tanay Kibe}
\email{tanayk@smail.iitm.ac.in}
\affiliation{Center for Quantum Information Theory of Matter and Spacetime, and Center for Strings, Gravitation and Cosmology, Department of Physics, Indian Institute of Technology Madras, Chennai 600036, India}
\author{Nehal Mittal}
\email{nehal.mittal@u-psud.fr}
\altaffiliation[Current affiliation: \,\,]{D\'{e}partement de Physique de l'\'{E}cole Normale Sup\'{e}rieure, 24 rue Lhomond, 75005 Paris, France}
\affiliation{Center for Quantum Information Theory of Matter and Spacetime, and Center for Strings, Gravitation and Cosmology, Department of Physics, Indian Institute of Technology Madras, Chennai 600036, India}
\author{Ayan Mukhopadhyay}
\email{ayan@physics.iitm.ac.in}
\affiliation{Center for Quantum Information Theory of Matter and Spacetime, and Center for Strings, Gravitation and Cosmology, Department of Physics, Indian Institute of Technology Madras, Chennai 600036, India}
\author{Pratik Roy}
\email{pratik@physics.iitm.ac.in}
\affiliation{Center for Quantum Information Theory of Matter and Spacetime, and Center for Strings, Gravitation and Cosmology, Department of Physics, Indian Institute of Technology Madras, Chennai 600036, India}

\date{\today}

\begin{abstract}
Investigating principles for storage of quantum information at finite temperature with minimal need for active error correction is an active area of research. We bear upon this question in two-dimensional holographic conformal field theories via the quantum null energy condition (QNEC) that we have shown earlier to implement the restrictions imposed by quantum thermodynamics on such many-body systems. We study an explicit encoding of a logical qubit into two similar chirally propagating excitations of finite von-Neumann entropy on a finite temperature background whose erasure can be implemented by an appropriate inhomogeneous and instantaneous energy-momentum inflow from an infinite energy memoryless bath due to which the system transits to a thermal state. Holographically, these fast erasure processes can be depicted by generalized AdS-Vaidya geometries described previously in which no assumption of specific form of bulk matter is needed. We show that the quantum null energy condition gives analytic results for the minimal finite temperature needed for the deletion which is larger than the initial background temperature in consistency with Landauer's principle. In particular, we find a simple expression for the minimum final temperature needed for the erasure of a large number of encoding qubits. We also find that if the encoding qubits are localized over an interval shorter than a specific \textit{localization length}, then the fast erasure process is impossible, and furthermore this localization length is the largest for an optimal amount of encoding qubits determined by the central charge. We estimate the optimal encoding qubits for realistic protection against fast erasure. We discuss possible generalizations of our study for novel constructions of fault-tolerant quantum gates operating at finite temperature.
\end{abstract}

\maketitle

\section{Introduction} 

Quantum thermodynamics is an emerging framework for the study of feasibility of quantum channels and resources like energy and entanglement needed for their function \cite{PhysRevLett.115.070503,Guryanova_2016,PhysRevE.93.022126,Yunger_Halpern_2016,Goold_2016,Gour_2018,RevModPhys.91.025001}. Most studies have been in the domain of finite-dimensional quantum systems where many thermodynamic notions have been generalized via information-theoretic quantities such as hypothesis-testing relative entropy between a state and a suitable Gibbs ensemble \cite{PhysRevE.93.022126}. Extensions of these studies to quantum field theories are of interest for finding general principles of construction of efficient quantum engines and fault-tolerant quantum gates using many-body systems. 

%cite{PhysRevA.73.012340,Brown:2014idi}

In the context of topologically ordered many-body systems, issues regarding fault tolerant quantum computation at finite temperature have been studied extensively \cite{KITAEV20032,Nayak:2008, Brennen:2008, pachos_2012}. Especially with regard to construction of \textit{self-correcting quantum memories} for storing quantum information without the need of active error correction \cite{RevModPhys.88.045005}, there are no-go results for one and two (spatial) dimensional models which utilize stabilizer codes (e.g. the toric code model), following simply from the result that the energy barrier between degenerate ground states does not scale with the system size \cite{Bravyi_2009,PhysRevLett.110.090502}. The lack of finite temperature stability of encoded information is supported by the rapid decay of topological order parameters at finite temperature \cite{Alicki:2009,Chesi:2010}. Quantum thermodynamics could be applied to design systems that are ideal for storing quantum memory. 
%{\pr Comment: Meaning of this sentence is not clear. We evade no-go theorems by going to a different system. Quantum thermodynamics does not change the status of no-go theorems where they are already known to hold. Maybe write: Studying quantum thermodynamics in different systems might help us learn how to evade the no-go theorems.}

Recently, it has been demonstrated in \cite{Kibe:2021qjy} that the quantum null energy condition (QNEC) \cite{Bousso_2016} can be a powerful tool to examine the restrictions, beyond classical thermodynamics, which physical processes should satisfy in two-dimensional conformal systems. For two-dimensional conformal field theories ($2D$ CFTs), QNEC states that the non-vanishing components of the energy-momentum tensor, namely $t_{\pm\pm}$ (with $\pm$ denoting the left and right moving future-directed null directions $x^\pm = t\pm x$), should be bounded from below by null derivatives of the entanglement entropy $S$ of any spatial interval ending at the point of observation as follows \cite{Wall:2011kb,Koeller_2016,Balakrishnan:2017bjg}:
\begin{equation}\label{Eq:QNECpm}
    \mathcal{Q}_\pm \coloneqq 2\pi \langle t_{\pm\pm} \rangle - S'' - \frac{6}{c}{S'}^2 \geq 0.
\end{equation}
Here, primes denote derivatives obtained from infinitesimal displacements of the endpoint coinciding with the point of observation along $\pm$ directions, and $c$ is the central charge of the CFT. Generic transitions between thermal states carrying momentum in holographic $2D$ CFTs driven by instantaneous energy-momentum inflow from an infinite  memoryless bath were examined in \cite{Kibe:2021qjy}. It was found that non-violation of QNEC not only implies that the entropy and temperature should increase as implied by classical thermodynamics, but also for a fixed increase in temperature (entropy) there are both lower and upper bounds on the increase in entropy (temperature). Furthermore, upper and lower bounds on the early time quadratic and late time ballistic growths of entanglement were obtained analytically. The strictest bounds implied by the inequality \eqref{Eq:QNECpm} are always produced by the semi-infinite entangling interval. These bounds thus could apply to any system whose infrared behavior is described by a CFT with a sufficiently large central charge and a sparse spectrum.

States which saturate \eqref{Eq:QNECpm} are termed quantum equilibrium states \cite{Ecker:2019ocp}. Essentially, these are states with Virasoro hair (chirally propagating excitations) on top of the vacuum or a thermal state. These are holographically dual to Ba\~nados geometries \cite{Banados:1998gg}. Methods developed in \cite{Kibe:2021qjy} allow the study of transitions between two such quantum equilibrium states in holographic systems. 
%(QNEC also implies that the bulk matter, which is determined just by the initial and final states, satisfies the classical null energy condition.)

In this work, we utilize these methods to study a simple encoding of a logical qubit in a quantum equilibrium state with two similar chirally propagating excitations on a finite temperature background, and its fast erasure in which the system transitions to a finite temperature state via instantaneous energy-momentum inflow from the bath. The non-violation of QNEC sets a lower bound on the final temperature in consistency with the Landauer principle \cite{Landauer:1961,sagawa2017second,Esposito_2010,Reeb_2014}, according to which information erasure should imply flow of energy from the information-carrying degrees of freedom to non-information-carrying degrees of freedom \cite{Reeb_2014}. For a large number of encoding qubits, the threshold final temperature needed for the erasure increases monotonically with the number of qubits, and has a simple analytic expression. Crucially, we show that the fast erasure is not allowed if the encoded information is localized on a length scale shorter than a specific value $b_c$ which is determined by the initial background temperature and the number of encoding qubits. The length scale $b_c$ turns out to be maximal when the number of encoding qubits are about $c/6$. We provide estimates for the number of encoding qubits etc for realistic applications to erasure tolerant quantum memories.

\begin{figure}[t]
\includegraphics[width=0.3\textwidth]{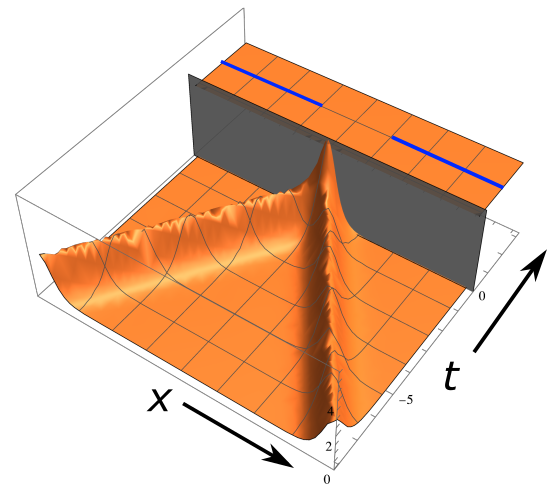}
\caption{A schematic depiction of the fast erasure process in which the energy density has been plotted as a function of the space and time coordinates. Two spacelike intervals (blue) that are related by reflection about $x=0$ have the same entanglement entropy.}
 %\caption{A plot of $\langle t_{\pm \pm} \rangle$ for the erasure protocol. The Lorentzian excitation exists for $t<0$ and is erased at $t=0$ such that the state for $t>0$ is thermal, with a constant stress tensor. Energy momentum injection from a bath occurs as a shock at $t=0$ and is indicated by the gray plane. Two semi-infinite spacelike intervals (blue) that are related by reflection about $x=0$ have the same von Neumann entropy.}
 \label{Fig:schematic}
\end{figure}
%In this paper we focus on transitions from inhomogeneous initial states, which have geometries with inhomogeneous traveling excitations  as their dual, to thermal final states, with $\mathcal{L}^{f}_{\pm} = \mu^{2}$, dual to Ba\~nados-Teitelboim-Zanelli (BTZ) black branes \cite{Ba\~nados:1992wn,Ba\~nados:1992gq}. See Fig. \ref{Fig:schematic}. The final states have the following (Hawking) temperature and thermodynamic entropy density
%\begin{equation}\label{Eq:T-s}
%T^f = \frac{1}{\pi} \mu, \quad s^{f} =\frac{c}{3} \mu.
%\end{equation}
%Transitions between thermal states  were studied in \cite{Kibe:2021qjy}. We follow the strategy presented there to compute the entanglement entropy and the QNEC (see supplementary material).
%The entanglement entropy of the reduced density matrix for a spacelike interval in the boundary CFT with end points $p_{1,2} = (x_{1,2},t_{1,2})$ for any state of the CFT can be holographically computed using the proper length $L_{\rm geo}$ of the bulk geodesic that is anchored to the end points at the regulated boundary $r = \frac{L^{2}}{\epsilon}$, where $\epsilon^{-1}$ corresponds to the ultraviolet energy cutoff in the dual CFT. The entanglement entropy is \cite{Ryu:2006bv,Hubeny:2007xt}
%\begin{align}
%    S_{\rm ent} = \frac{c}{6}\frac{L_{\rm geo}}{L}.
%\end{align}
%QNEC involves variation of the entanglement entropy under null deformations of one of the end points $p_{1,2}$.

\section{Holographic quantum memory\\ and fast erasure}
A simple encoding of a logical qubit into the Hilbert subspace of a CFT spanned by the Virasoro descendents of the vacuum, which can readily be generalized to a finite temperature background, is as follows. Consider the stereographic projection of the point on the Bloch sphere representing the logical qubit state onto a point $z_*$ on the complex plane. The Lorentzian continuation describes a point with coordinates $(t_*, x_*)$ or equivalently $(x^+_*,x^-_*)$. The encoding is achieved by the state with $\langle t_{\pm\pm} \rangle =\mathcal{L}_\pm(x^\pm) = \mathcal{L}(x^\pm - x^\pm_*)$ comprising two similar left and right moving energy-momentum excitations on vacuum or a finite temperature background. The logical qubit can be readily decoded by measuring the energy density at any time and deducing the coordinates $(t_*, x_*)$ where the centers of the two chirally propagating excitations coincide. For simplicity, we consider $(t_*, x_*)$ to be the origin.

\begin{figure}[t]
\includegraphics[width=0.35\textwidth]{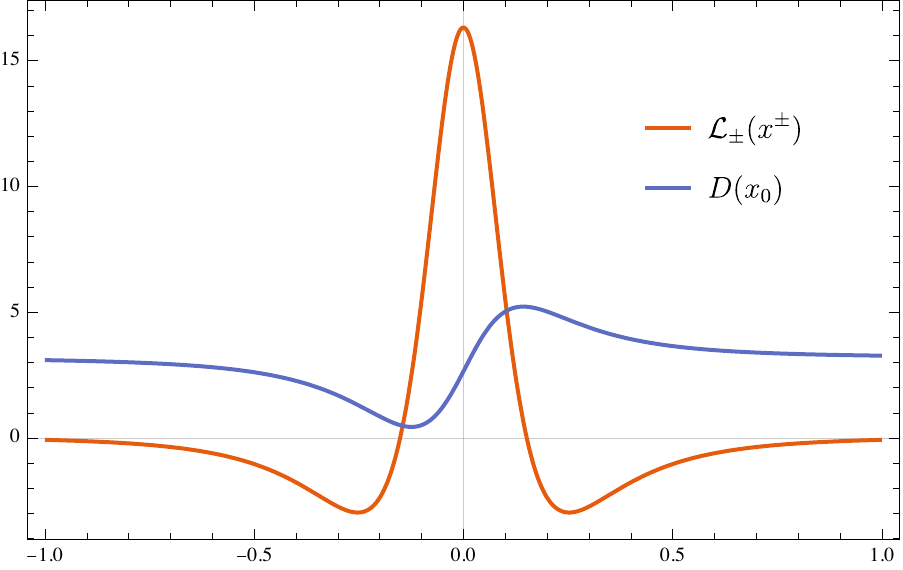}
\caption{A plot of $\mathcal{L}(x^{\pm})$ (or equivalently energy density at the moment of quench $t=0$) and $D(x_{0})$, the coefficient of quadratic growth of entanglement entropy of the interval $x\geq x_0$. Here $a=0.01$, $b=0.19$, $T^{i} = \frac{0.2}{\pi}$ and $T^{f} = \frac{1}{\pi}$. }\label{Fig:lump}
\end{figure}

More specifically, we will consider $\mathcal{L}(x^{\pm})$ of the form
\begin{align}           \label{Eq:Lansatz}
   \mathcal{L}(x^{\pm}) = -\frac{1}{2} \text{Sch}\left(\int^{x^{\pm}} w(y) dy,\,x^{\pm}  \right) +w(x^{\pm})^{2}
\end{align}
where Sch denotes the Schwarzian derivative, and $w(x)$ is a positive constant plus any smooth function with finite support such that $w(x) > 0$. As a specific example, we choose
\begin{equation}\label{Eq:w}
    w(x) = \frac{a}{b^{2} +x^{2}} + d,
\end{equation}
which gives
\begin{align}
\label{Eq:Lpm}
    \mathcal{L}(x^{\pm}) =&\, \frac{3 a^2 {x^{\pm}}^2+\left(a+d \left(b^2+{x^{\pm}}^2\right)\right)^4}{\left(b^2+{x^{\pm}}^2\right)^2 \left(a+d \left(b^2+{x^{\pm}}^2\right)\right)^2}  \\ 
     +&\, \frac{a \left(b^2-3 {x^{\pm}}^2\right) \left(a+d \left(b^2+{x^{\pm}}^2\right)\right)}{\left(b^2+{x^{\pm}}^2\right)^2 \left(a+d \left(b^2+{x^{\pm}}^2\right)\right)^2}\nonumber ,
\end{align}
where $a$ is the amplitude and $b$ the width of the Lorentzian function $w$. See Fig.\! \ref{Fig:lump} for a plot of $\mathcal{L}(x^{\pm})$.  Since $\lim_{x^{\pm} \to \pm \infty} \mathcal{L}_{\pm}(x^{\pm}) = d^{2}$, $\mathcal{L}(x^\pm)$ is described by two similar chirally propagating excitations on top of a background temperature  $T = \frac{d}{\pi}$. In a holographic CFT, with large central charge $c$ and sparse spectrum, this state can be described by a Ba\~nados geometry (see below). 

%It is useful to define a measure of the amount of information that can be encoded in the initial excitation, and the von Neumann entropy of the initial state is a natural choice \cite{quantumcoding}. 
The number of qubits used for the encoding can be estimated from the von-Neumann entropy that is in excess of the background thermal contribution. Using the holographic entanglement entropy prescription \cite{Ryu:2006bv,Hubeny:2007xt}, one can readily obtain that the entanglement entropy of an interval with endpoints $(-\frac{l}{2}, t)$ and $(\frac{l}{2},t)$ in the large $l$ limit is
\begin{align}\label{Eq:sinfo}
    S_{\rm ent} &=  S_{\rm ent}^{\rm th}(d) + s_{\rm info} + \mathcal{O}\left(\frac{1}{l}\right),&
    s_{\rm info} &= \frac{c}{6}\frac{2 \pi a}{b},
\end{align}
where $S_{\rm ent}^{\rm th} (d) = \frac{c}{3}\left(d l - \log (2 d) \right)$ is simply the thermal contribution corresponding to the temperature $T= \frac{d}{\pi}$. Therefore, $s_{\rm info}$, which is independent of the entangling length $l$, provides an estimate of the number of encoding qubits.

We model information erasure as a quench that transforms the inhomogeneous state with encoded information to a thermal state with $\mathcal{L}_{\pm} = \mu^{2}$ (and with temperature $\mu/\pi$) as depicted in Fig. \ref{Fig:schematic}. The quench involves inhomogeneous inflow of energy and momentum from an infinite memoryless bath and can be realized physically via appropriate time-dependent couplings between the bath and the system \cite{Balasubramanian:2010ce,Hubeny:2010ry,Chesler:2013lia}. In the holographic setup, we can model this via an appropriate bulk energy-momentum tensor which is infalling from the boundary of spacetime without assuming any explicit realization of the bulk matter. We will consider the energy-momentum inflow from the bath describing the quench to be instantaneous \footnote{The time-scale of the quench should be much larger than the ultraviolet cutoff of the CFT.}. In this case, the form of the bulk energy-momentum tensor is fixed completely by initial and final states via the Israel junction conditions \cite{Israel:1966rt}. To simplify the discussion, we assume that the quench occurs at $t=0$, when the propagating excitations coincide. However, our results remain valid even if the quench occurs before or after.

The holographic description of the full process \cite{Brown:1986nw,Maldacena:1997re,Henningson:1998gx,Balasubramanian:1999re} in the limit of large central charge is given by a generalized $2+1$ dimensional AdS-Vaidya geometry involving a transition between the Ba\~nados geometry depicting the initial encoding state to a BTZ (Ba\~nados-Teitelboim-Zanelli) black hole \cite{Banados:1992wn,Banados:1992gq} describing the final thermal state as described in \cite{Kibe:2021qjy}. The metric, which solves Einstein's equations with a negative cosmological constant $\Lambda = - 1/L^2$, takes the form
\begin{eqnarray}\label{Eq:metric}
{\rm d}s^2 &=& 2 {\rm d}r {\rm d}t -\left(\frac{r^2}{L^2} - 2 m(t,x)L^2\right) {\rm d}t^2  +\frac{r^2}{L^2} {\rm d}x^2,
\end{eqnarray}
which is a special case of the more general transitions between arbitrary Ba\~nados geometries considered in \cite{Kibe:2021qjy}. The dual CFT shares the coordinates $t,x$, and lives at the boundary $r \to \infty$ of this spacetime. The metric is supported by a bulk energy momentum tensor $T_{\mu \nu}$, which is conserved and traceless in the background metric  \eqref{Eq:metric}, and whose only non-vanishing component is
\begin{eqnarray}\label{Eq:Tbulk}
T_{tt} &=& \frac{q(t,x) L^2}{r} .
%+ \frac{\partial_x p(t,x) L^4}{r^2} + \frac{p(t,x) j(t,x)L^6}{r^3},\nonumber\\
%\,\, T_{tx} &=& \frac{p(t,x)L^2}{r}.
\end{eqnarray}
For our specific fast erasure process, we have
\begin{eqnarray}\label{Eq:M-J}
m(t,x) &=& \theta(-t)(\mathcal{L} (x^+) + \mathcal{L}(x^-)) 
+ 2\theta(t) \mu^2,
%, \nonumber\\
%j &=& 0.
\end{eqnarray}
with $\mathcal{L}(x)$ given by the even function \eqref{Eq:Lpm} and
\begin{eqnarray}\label{Eq:q-p}
8\pi G q(t,x) &=& 2 \delta(t)\left(\mu^2- \mathcal{L} (x)\right),
%\nonumber\\ 
%8\pi G p &=& \delta(t)(\mathcal{L}^f_+ (x) - \mathcal{L}^i_+ (x) -\mathcal{L}^f_- (-x) + \mathcal{L}^i_- (-x)).
\end{eqnarray}
with $G$ being the three dimensional Newton's constant. The bulk matter \eqref{Eq:Tbulk} is thus an inhomogeneous infalling null shell. The non-violation of QNEC, as described below, requires $q(t,x) > 0$ due to which the bulk matter satisfies the classical null energy condition . 

The energy-momentum tensor of the dual CFT state can be be obtained via the procedure of holographic renormalization \cite{Henningson:1998gx,Balasubramanian:1999re}. The non-vanishing components are
\begin{equation}
\langle t_{\pm \pm} \rangle =\frac{c}{12 \pi}\left(\mathcal{L}(x^\pm)\theta(-t) + \mu^2 \, \theta(t)\right).
\end{equation}
where $c = 3L/(2G)$ is the central charge \cite{Brown:1986nw,Henningson:1998gx,Balasubramanian:1999re}. The Ward identity $\partial_\mu \langle t^{\mu\nu} \rangle = f^\nu$, where $f^\nu =L (q(t,x), 0)$ is implied by \eqref{Eq:q-p}, so that $q(t,x)$ describes the heat inflow from the bath to the CFT. The condition $q(t,x)>0$, implied by the non-violation of QNEC as discussed above, is thus equivalent to the Landauer principle which implies that energy should flow irreversibly from information carrying degrees of freedom to non-information carrying degrees of freedom during information erasure. 

The time-dependent entanglement entropy is given by the holographic entanglement entropy prescription \cite{Ryu:2006bv,Hubeny:2007xt}, and therefore $\mathcal{Q}_\pm$ in \eqref{Eq:QNECpm} can be computed \textit{analytically} following the method developed in \cite{Kibe:2021qjy} via the use of uniformization maps. For a summary, see the Supplemental Material.
%The latter utilizes two separate uniformization maps which takes the initial Ba\~nados geometry and the final BTZ black hole respectively to the Poincar\'{e} patch geometry (given by the metric \eqref{Eq:metric} with $m(t,x) =0$) representing the vacuum state. The hypersurface $t=0$ corresponding to the quench have two separate images in these  Poincar\'{e} patches which are glued such that the physical coordinates $r$ and $x$ describing the original metric \eqref{Eq:metric} are continuous. These allows an explicit analytic computation of the length of the bulk geodesics anchored to the boundary points, and hence the entanglement entropy. For a summary, see the Supplemental Material.

%Our results don't qualitatively change if the quench occurs at $t \neq 0$. QNEC is then used as a tool to place constraints on such an erasure. Note that in 2d CFTs chiral excitations live forever and don't thermalize. The only way to thermalize such excitations is via coupling to a bath. Therefore it is valid to refer to such a \textit{forced} thermalization as information erasure. It is natural to ask if the kind of excitations described here can be created from the vacuum or a thermal state via a quench. We find that QNEC disallows such a creation protocol. We also find that QNEC disallows the erasure of infinitely de-localized periodic excitations via such a quench. See supplementary material for details on both.

Let us discuss first how the entanglement entropy evolves. It is to be noted that although the bulk metric becomes that of the BTZ black hole and the energy-momentum tensor becomes thermal immediately after quench, the state does not thermalize immediately as is evident from the time-dependence of the entanglement entropy of an interval of length $l$. It has been shown in \cite{Kibe:2021qjy} that the thermalization time for an arbitrary momentum carrying thermal state is $l/2$ supporting earlier results \cite{Hubeny:2013hz,Liu:2013iza,Liu:2013qca}. 

As shown in Fig. \ref{Fig:schematic}, owing to reflection symmetry $x\rightarrow-x$ of our geometry, the entanglement entropy of two spatial intervals related by reflection about origin should be identical. Also $\mathcal{Q}_+$ of one of these intervals would be same as $\mathcal{Q}_-$ of the other for displacements of the endpoint closer to the origin. 

Immediately after quench, i.e. at $t=0^+$, the entanglement entropy $S(t)$ of a semi-infinite interval $x\geq x_0$ grows quadratically, $S(t) \approx S_0+  D t^{2}$, with
\begin{equation}
   D = \mu \left(4 \mu -2 d -\frac{w'(x_{0})}{w(x_{0})}-2 w(x_{0})\right).
   \label{Eq:Diff}
\end{equation}
At larger times, the entanglement entropy of these semi-infinite intervals grows linearly as $S(t) \approx 2 (s^{f} - s^{i}) t$ with $s^f = (c/3) d$ and $s^i = (c/3) \mu$ being the initial and final thermal entropy densities as observed before in \cite{Kibe:2021qjy}. This is consistent with the light-cone like spreading of entanglement seen in \cite{Calabrese:2009qy,Liu:2013qca,Liu:2013iza,Calabrese:2016xau,Cheneau2012,PhysRevLett.111.127205,Kaufman:2016} from both ends of an interval.

%also exists in such inhomogeneous quenches. These results extend those seen for instantaneous transitions between thermal states \cite{Liu:2013iza, Liu:2013qca,Hubeny:2013hz,Kibe:2021qjy}.

The quadratic growth coefficient $D$ in \eqref{Eq:Diff} is plotted in Fig. \ref{Fig:lump} as a function of the endpoint $x_0$ of the semi-infinite interval. Note that the reflection asymmetry originates from our choice of interval $x\geq x_{0}$. We note that the coefficient $D$ is not positive definite and demanding $D \geq 0$ already bounds the final temperature $T^f$ from below. This bound can be easily evaluated when the amplitude ($\propto a$) of the chiral excitations is large for fixed width $b$ and initial background temperature $T^i = d/\pi$. In this limit, the number of encoding qubits $s_{\rm info}$ given by \eqref{Eq:sinfo} is large, and we find that $D \geq 0$ implies
\begin{equation}
    T^f  \geq T^{i} +\frac{a}{2 \pi b^{2}} = T^{i} +\frac{3 s_{\rm info}}{2 c\pi^2 b}.
    \label{Eq:diffbound}
\end{equation}
QNEC non-violation places a stronger lower bound than positivity of $D$ as reported below.

\section{Erasure tolerance}

The QNEC inequalities \eqref{Eq:QNECpm} produce the strongest bounds for the erasure process when we consider the entangling region to be a semi-infinite interval as was found previously \cite{Kibe:2021qjy} in the context of transitions between thermal states (see Supplementary Material for details). Owing to the reflection symmetry of the setup, we consider half line intervals $x \geq x_{0}$ in what follows unless explicitly mentioned otherwise.
%There is no loss of generality  since $\mathcal{Q}_{\pm}(x>x_{0}) = \mathcal{Q}_{\mp}(x<x_{0})$. 
As detailed in the Supplementary Material, QNEC prohibits the fast erasure process completely when the initial background temperature ($d/\pi$) is zero and also if any encoding is done via an $\mathcal{L}(x)$ which has infinite extent (and entropy). The interesting question however is the minimal value of the final temperature ($T_c = \mu_c/\pi$) required by the erasure if the encoding is done by a finite number of qubits on a finite (initial) temperature background. The results obtained from QNEC non-violation are consistent with the Landauer principle which demands that information erasure must be accompanied by increase in the thermodynamic entropy of the non-information-bearing (thermal) degrees of freedom.

%For $T^i = 0$, we find remarkably that $\mathcal{Q}_{+}$ is always violated (see supplementary material), so that (instantaneous) information erasure is possible only if $T^i \neq 0$.
%Generally, QNEC is satisfied provided $T^f - T^i \geq T_{c}$, where $T_c$ depends on the function $w(x)$. Thus QNEC demands that information erasure from the information bearing degrees of freedom must be accompanied by a minimum increase in the thermodynamic entropy of the non-information bearing degrees of freedom (which here is the thermal background).

For a large number of encoding qubits, $s_{\rm info}$, a fixed background temperature ($T^i = d/\pi$)  and a fixed width ($b$) of the encoding excitation, QNEC non-violation implies
\begin{equation}
\label{Eq:qnecbound1}
    T^{f}  \geq  T^{i} +\frac{a}{\pi b^{2}}= T^{i} +\frac{3 s_{\rm info}}{c\pi^2 b} =T^{i} + \frac{1}{\pi}\left(\max w(x) \right).
\end{equation}
Note that this is stronger than the bound \eqref{Eq:diffbound} obtained by demanding the increase of entanglement entropy just after quench. Interestingly, we also find the same bound for large $a$ and $b$ but with fixed $s_{\rm info}$. The bound stated in terms of $w(x)$ also holds for $w(x) = a\, \text{sech}(x/b) + d $ in both these limits indicating that it could be valid for any encoding excitation with a finite support.
%{\pr might} provide\sout{s} a universal lower bound on the final temperature for erasure in terms of the \sout{uniformization} {\pr encoding} map for any encoding via finite support excitations.
\begin{figure}[t]
\includegraphics[width=0.35\textwidth]{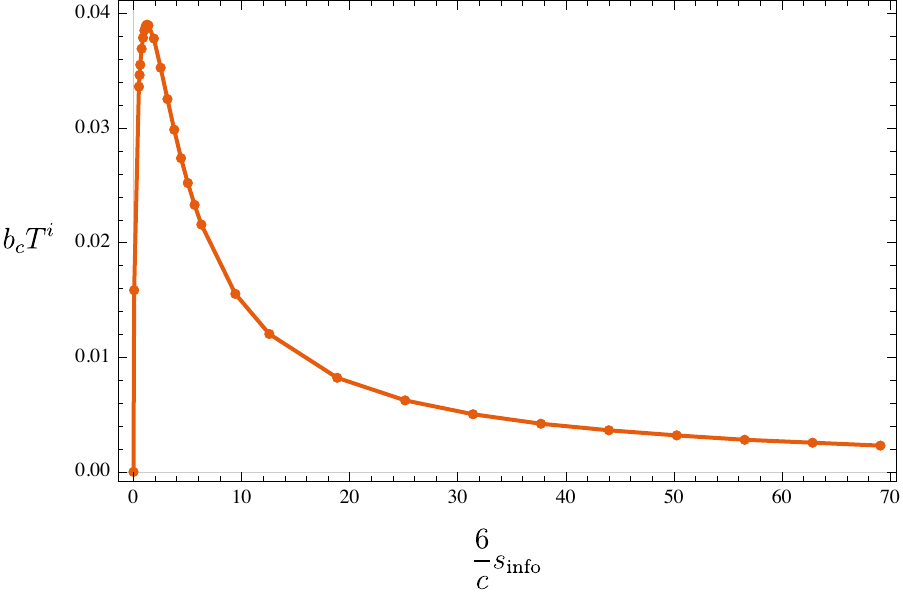}
\caption{The critical localization length $b_{c}$ is plotted against $s_{\rm info}$ for a half line interval and fixed $d=0.2$. $b_{c} \to 0$ as $s_{\rm info} \to 0$, since the excitation disappears in this limit.}\label{Fig:localization}
\end{figure}

Remarkably, if one squeezes the encoding qubits to a width $b < b_c$ for a fixed background temperature $T^i =d/\pi$, the fast erasure process is disallowed as the minimal final temperature needed for the erasure diverges. For widths $b \geq b_c$, one can always satisfy QNEC inequalities if $T^f \geq T^c$ making erasure possible. When $b< b_c$, $\mathcal{Q_+}<0$ for intervals $x\geq x_0$ (and $\mathcal{Q_-}<0$ for intervals $x\leq x_0$) violating QNEC for any final temperature. In fact, in this case, for large final temperature $\mu/\pi$, $\mathcal{Q}_+(t)/\mu \approx f(t)$, with $f(t)$ independent of $\mu$ for any interval. This feature leads to a simple algorithm for obtaining $b_c$ for a fixed $s_{\rm info}$ (see Supplementary Material). We find that both $b_c$ and $T_c$ increase with the length of the entangling interval monotonically, and reach finite values for the semi-infinite interval which thus puts maximum restriction on the parameter space where erasure is possible.

The dimensionless critical localization length $b_c T_i$ (for the semi-infinite interval) depends on $s_{\rm info}$ as shown in Fig.\! \ref{Fig:localization}. We note as $s_{\rm info}\rightarrow\infty$, $b_{c} \sim 0.4/s_{\rm info}$. This is consistent with \eqref{Eq:qnecbound1} which states that in this limit erasure is possible for sufficiently large $T^f$ for any width $b$. Similarly in the limit $s_{\rm info}\rightarrow0$ erasure should be possible for any width $b$ implying that $b_cT^i$ should vanish in this limit also. As shown in Fig.\! \ref{Fig:localization}, the dependence of $b_c T_i$ on $s_{\rm info}$ is non-monotonic and $b_c T^i$ obtains its maximal value $\approx 0.04/ T^i$ for an optimal $s_{\rm info}\approx c/6$ allowing protection against fast erasure with reasonably small density of encoding.

\begin{figure}[t]
\includegraphics[width=0.35\textwidth]{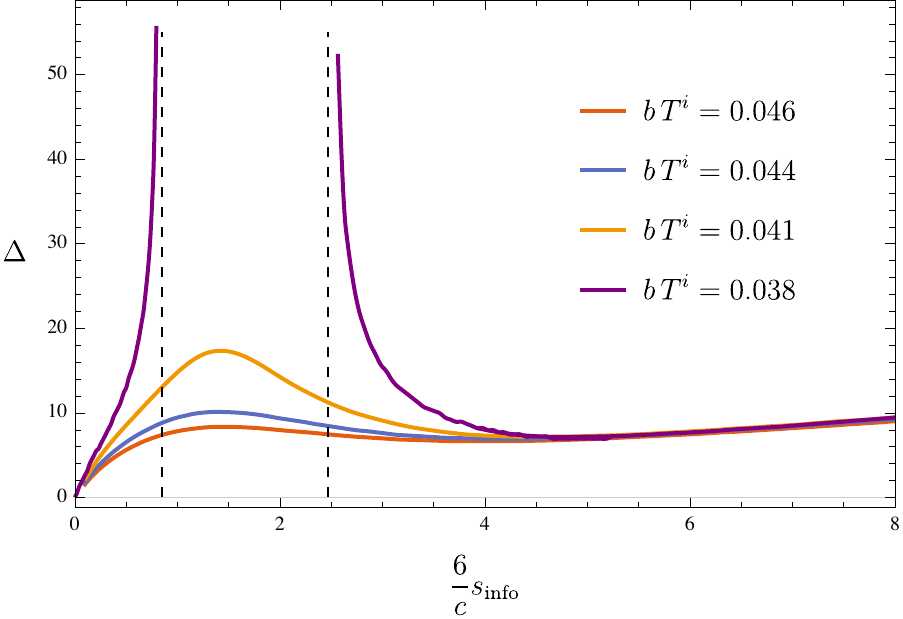}
 \caption {The minimum value of $(T^{f}-T^{i})/T^i$ necessary for the erasure process ($\Delta$) is plotted against the number of encoding qubits $s_{\rm info}$ for various fixed encoding widths $b$ when background temperature $T^i = 0.2/\pi$. For large $s_{\rm info}$, we obtain \eqref{Eq:qnecbound1}.}\label{Fig:nonmon}
\end{figure}

In Fig. \ref{Fig:nonmon}, $\Delta$, the minimal value of $(T^f- T^i)/T^i$ needed for the erasure, has been plotted as a function of $s_{\rm info}$ for fixed widths $b$ and a background temperature $T^i$ of the encoding. For large $s_{\rm info}$, the dependence is given by \eqref{Eq:qnecbound1}. However, as expected from the above discussion, the dependence is prominently non-monotonic for intermediate values of $s_{\rm info}$ especially for small values of $b$. In fact, as expected from Fig. \ref{Fig:localization} and shown explicitly in Fig. \ref{Fig:nonmon}, for $bT^i \lesssim 0.04$, there are two values of $s_{\rm info}$, which we denote as $s_{\rm info}^1$ and  $s_{\rm info}^2$, such that $\Delta$ diverges as $1/(s_{\rm info}- s_{\rm info}^{1,2})$ as $s_{\rm info}$ approaches $s_{\rm info}^1$ and $s_{\rm info}^2$ from below and above respectively, implying protection against erasure for $s_{\rm info}^1 \leq s_{\rm info} \leq s_{\rm info}^2$. 

%Furthermore, the region protected against erasure increases in size as we decrease $b$ further. This indicates that for a given amount of initial information $s_{\rm info}$ and fixed $d$ there exists a critical localization length $b_{c}$ such that if information is localized below $b=b_{c}$ it is protected against fast erasure, i.e., QNEC is violated for arbitrarily large final temperature. The variation of this critical localization length with $s_{\rm info}$ is shown in fig.\! \ref{Fig:localization}. At large $s_{\rm info}$ the inequality \eqref{Eq:qnecbound1} should hold and, for arbitrarily small $b$, one can find a large enough $T^f$ such that QNEC is satisfied. Thus $b_{c}\to0$ as $s_{\rm info} \to \infty$ in fig.\! \ref{Fig:localization}. In particular, for $s_{\rm info} \to \infty$ we find that $b_{c} \sim \frac{0.4}{s_{\rm info}}$, whereas at intermediate amounts of information we see a maximal localization length. 
%For a given $s_{\rm info}$, $b_{c}$ increases as the length of the entangling interval is increased and it is maximum for the half line. The minimal value of $\Delta T_{H}$ in fig.\! \ref{Fig:nonmon} increases as the length of the entangling interval increases and is maximum for the half line. Thus the bounds shown in fig.\! \ref{Fig:nonmon} and \ref{Fig:localization} are the strictest.

So far we have considered arbitrarily large final temperatures. However, the final temperature in an experimental setup will be only as large as that of the environment. This fixed value of maximal final temperature and a fixed $s_{\rm info}$ determine a point in Fig.\! \ref{Fig:nonmon}. Then there exists $b = b^{*}$ such that the corresponding curve $\Delta$ vs $s_{\rm info}$ passes through this point. In the experimental setup, erasure protection is then achieved for $b \leq b_*$ since then one requires a value of $\Delta$ for erasure which is higher than what the environment can provide. If the background temperature $T^{i}$ of the quantum memory is $50 K$, and the environment provides $T^f = 300 K$, then for $c = 100$ we find that to protect 50 encoding qubits we need the excitations to be localized to $\lesssim 2 \mu$m. For realistic applications, $T^f$ needs to be smaller than the microscopic energy scale at which the CFT ceases to be a valid description.

\section{Discussion}
It should be interesting to study whether squeezing the encoding qubits protects against other errors, especially those corresponding to shifts of the center of $\mathcal{L}(x)$, or distortions which produce multiple peaks and therefore make the decoding erroneous. Furthermore, one should also investigate slower erasure processes in which energy and momentum are pumped in from the bath over a long time scale. We would also like to study implementation of quantum gates via unitary conformal transformations on the encoded qubits and study their fault tolerant properties. One would expect that one can protect against different errors simultaneously by exploiting the infinite dimensional Virasoro group suitably.

%\lipsum[1-4]

 \begin{acknowledgments}
It is a pleasure to thank Souvik Banerjee, Abhishek Chowdhury, Arnab Kundu, Prabha Mandayam, Pramod Padmanabhan and Marios Petropoulos for helpful discussions. The research of AB is supported by IFCPAR/CEFIPRA project 6304-3 and the research of TK is supported by the Prime Minister's Research Fellowship (PMRF). AM acknowledges the support of the Ramanujan Fellowship of the Science and Engineering Board (SERB) of the Department of Science and Technology (DST) of India, IFCPAR/CEFIPRA project 6304-3, the new faculty seed grant of IIT Madras and the additional support from the Institute of Eminence scheme of IIT Madras funded by the Ministry of Education of India. NM would like to thank IIT Madras and his home institutions Paris-Saclay University and École Normale Supérieure for their hospitality.
\end{acknowledgments}

\newpage
\section*{Supplementary Material}
\subsection*{Cut and glue method for entanglement entropy and QNEC}
%While calculating geodesic lengths in arbitrary Ba\~nados geometries (dual to quantum equilibrium states) is challenging, this can be readily done in the Poincar\'{e} patch. 
A Ba\~nados geometry is dual to a quantum equilibrium state \cite{Ecker:2019ocp} for which 
\begin{equation}
    \langle t_{\pm \pm}\rangle = \frac{c}{12\pi}\mathcal{L}_\pm(x^\pm)
\end{equation}
and is described by the metric 
\begin{eqnarray}\label{Eq:metricB}
{\rm d}s^2 &=& 2 {\rm d}r {\rm d}t -\left(\frac{r^2}{L^2} - 2 (\mathcal{L}_+(x^+)+\mathcal{L}_-(x^-))L^2\right) {\rm d}t^2 \nonumber\\&&+ 2 (\mathcal{L}_+(x^+)-\mathcal{L}_-(x^-)) L^2{\rm d}t {\rm d}x  +\frac{r^2}{L^2} {\rm d}x^2.
\end{eqnarray}
Any Ba\~nados geometry can be uniformized to the Poincar\'{e} patch, given by the above metric with $\mathcal{L}_\pm(x^\pm) =0$ and dual to the vacuum state, using the following uniformization map \cite{Kibe:2021qjy}
\begin{eqnarray}\label{Eq:uniformization}
T &=& \frac{1}{2}\Bigg(X_b^+(x^+) +X_b^-(x^-)\nonumber\\ &&+ \frac{{X_b^+}'(x^+) +{X_b^-}'(x^-)-2 \sqrt{{X_b^+}'(x^+){X_b^-}'(x^-)}}{\frac{r}{L^2}-\frac{{X_b^+}''(x^+)}{2{X_b^+}'(x^+)}-\frac{{X_b^-}''(x^-)}{2{X_b^-}'(x^-)}}\Bigg),\nonumber\\
X &=& \frac{1}{2}\Bigg(X_b^+(x^+) -X_b^-(x^-)\nonumber\\&&+ \frac{{X_b^+}'(x^+) -{X_b^-}'(x^-)}{\frac{r}{L^2}-\frac{{X_b^+}''(x^+)}{2{X_b^+}'(x^+)}-\frac{{X_b^-}''(x^-)}{2{X_b^-}'(x^-)}}\Bigg),\nonumber\\
R &=& L^2\frac{{\frac{r}{L^2 }-\frac{{X_b^+}''(x^+)}{2{X_b^+}'(x^+)}-\frac{{X_b^-}''(x^-)}{2{X_b^-}'(x^-)}}}{\sqrt{{X_b^+}'(x^+){X_b^-}'(x^-)}},
\end{eqnarray}
where $(T, R, X)$ denote the Poincar\'{e} patch coordinates. Clearly $X_{b}^{\pm} (x^{\pm})$ are the  boundary lightcone coordinates of the Poincar\'{e} patch as evident from the limit $r\rightarrow\infty$. For the uniformization map to be valid, $X_{b}^{\pm} (x^{\pm})$ should satisfy
\begin{equation}\label{Eq:Sch}
   {\rm Sch}( X_b^\pm(x^\pm), x^\pm) = - 2 \mathcal{L}_\pm(x^\pm).
\end{equation}
This equation can be solved by first solving the Hill equation
\begin{equation}
    	\Psi_\pm'' (x^{\pm})- \mathcal{L}_{\pm}(x^\pm) \Psi_\pm(x^\pm) = 0,
\end{equation}
 which has two linearly independent solutions $\Psi^{1}_\pm$ and $\Psi^{2}_\pm$.
One can then readily verify that
\begin{equation}
    X_{b}^\pm(x^\pm) = \frac{\Psi^{1}_\pm(x^\pm)}{\Psi^{2}_\pm(x^\pm)}
\end{equation}
is a solution to \eqref{Eq:Sch}. Solutions for $X_{b}^{\pm}$ can be analytically generated as follows. For $\mathcal{L}_\pm(x^\pm) = 1$ it is easy to see that
\begin{equation}
   \Psi^1_\pm(x^\pm) = \frac{1}{\sqrt{2}}\sinh x^\pm, \quad \Psi^2_\pm (x^\pm) = \frac{1}{\sqrt{2}}\cosh x^\pm
\end{equation}
are two independent solutions to the Hill equation. 
%The conformal covariance of the Hill equation then relates these solutions to the solutions corresponding to $\mathcal{L}_\pm$ used in the main text. Under the transformation $x^\pm \to \int^{x^\pm} w(y) dy$ the solutions $\Psi_{1,2}$ and $\mathcal{L}_\pm$ transform as
The conformal covariance of the Hill equation then implies that under the transformation $x^\pm \to \int^{x^\pm} w(y) dy$ (for any positive definite function $w(y)$), 
\begin{align}
\mathcal{L}_\pm(x^\pm)& \to -\frac{1}{2} {\rm Sch}\left(\int^{x^\pm} w(y) dy ,x^\pm\right) + w(x^\pm)^{2},\\
    \Psi^{1}_\pm(x^\pm) &\to \frac{1}{\sqrt{2 w(x^\pm)}} \sinh\left(\int^{x^\pm} w(y) dy\right),\\
    \Psi^{2}_\pm(x^\pm) &\to \frac{1}{\sqrt{2 w(x^\pm)}} \cosh\left(\int^{x^\pm} w(y) dy\right).
\end{align}
For the transformation of $\mathcal{L}_\pm(x^\pm)$ above we have added the anomalous term with appropriate central charge. We can thus obtain the uniformization map corresponding to the Lorentzian $w(x)$ described in the main text.

\begin{figure}[t]
\includegraphics[width=0.4\textwidth]{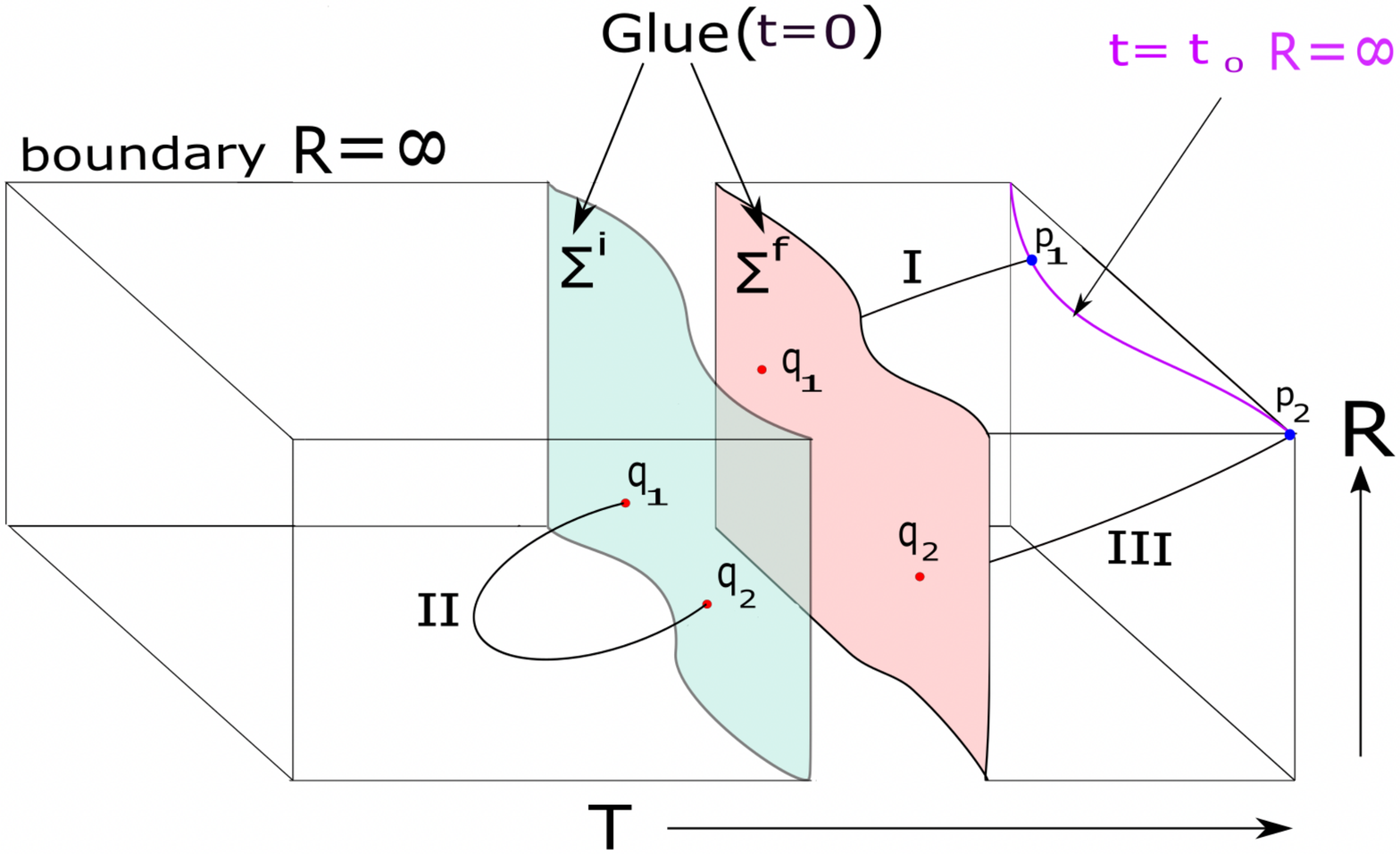}
\caption{Schematic diagram of the cut and glue method to compute the entanglement entropy. The left and right halves correspond to the two  Poincar\'{e} patches before and after the quench at $t=0$. The two glueing hypersurfaces $\Sigma^{i,f}$ are coloured. Points on these hypersurfaces with the same $x,r$ coordinates are identified. A geodesic ending at the boundary points $p_{1,2}$ is shown (black solid curve).}\label{Fig:Glue}
\end{figure}

We note that the metric \eqref{Eq:metric} is essentially two Ba\~nados geometries separated by an inhomogeneous null shell infalling from the boundary at the time of the quench $t=0$ and satisfying Israel junction conditions \cite{Kibe:2021qjy}. In order to compute the entropy and QNEC, we use two different uniformization maps \eqref{Eq:uniformization} for the two respective Ba\~nados geometries before and after the quench. The hypersurface $t=0$ has two separate images $\Sigma^{i,f}$ in the respective Poincar\'{e} patches  which are glued together by identifying points with the same \textit{physical} coordinates $x$ and $r$ as shown in Fig. \ref{Fig:Glue}. 

Geodesic lengths can be readily computed in the Poincar\'{e} patch in terms of the endpoints $(T_{1,2}, X_{1,2}, R_{1,2})$ using the formula
\begin{equation}
    L_{\rm geo} = L\ln (\xi + \sqrt{\xi^2 -1}),
\end{equation}
with $\xi = (Z_1^2 + Z_2^2 - (T_1 +Z_1 - T_2 -Z_2)^2 + (X_1 - X_2)^2)/2 Z_1 Z_2$ and $Z_{1,2} = L^2/R_{1,2}$.
For transitions between Ba\~nados geometries, we can thus use the uniformization map to analytically calculate geodesic lengths in the two Poincar\'e patches and sum the contributions. Fig. \ref{Fig:Glue} shows a geodesic in such a geometry cut into three arcs by the gluing hypersurface. The points $q_{1,2}$ where the geodesic intersects the gluing hypersurface can be analytically obtained if the $t>0$ geometry is a non-rotating BTZ spacetime. The lengths of each of these segments can therefore be readily computed using the geodetic distance formula in the corresponding Poincar\'{e} patch. Similarly the variation of the entanglement entropy under null deformations of one of the end points $p_{1,2}$ can be computed and the QNEC can be obtained (see \cite{Kibe:2021qjy} for details).

\subsection*{The $l \to \infty$ limit gives the strongest QNEC inequality}
\begin{figure}[t]
\includegraphics[scale=.6]{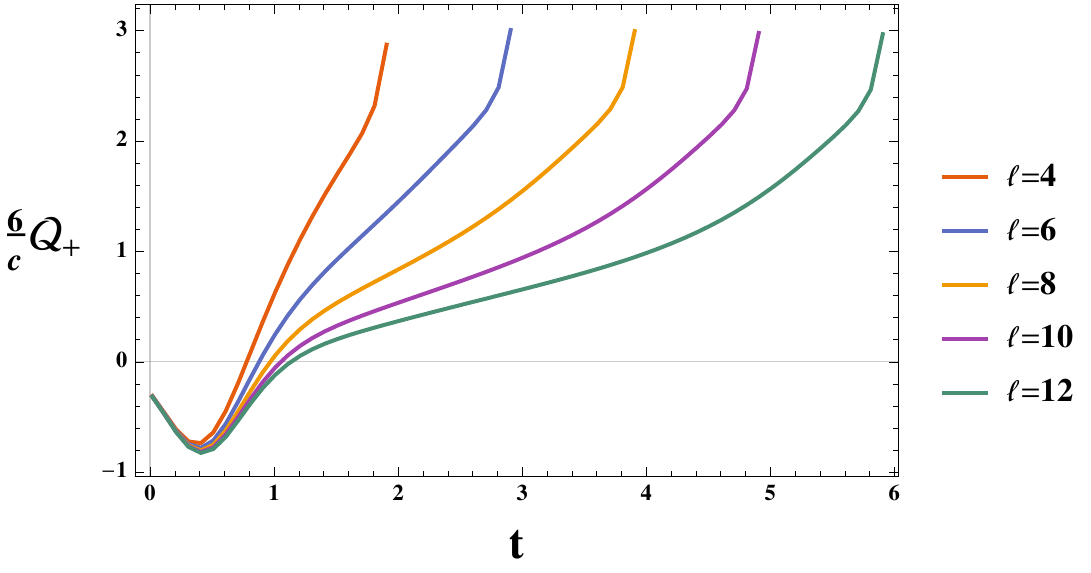}\,
\includegraphics[scale=.6]{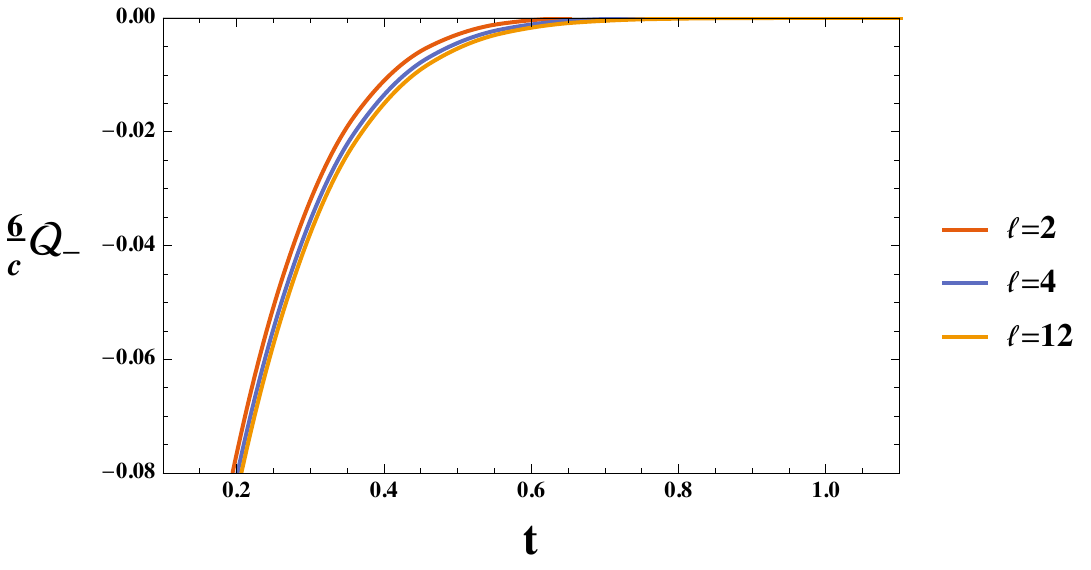}
\caption{$\mathcal{Q}_+$ (top) is plotted for $0< t < l/2$ and $\mathcal{Q}_-$ (bottom) is plotted for $ 0< t < {\rm min}(1.1, l/2)$ for various lengths $l$ of the entangling interval and the fast quench that transforms a Lorentzian excitation with $a = 1$, $b=1$ and $d=0.1$ to a thermal state with temperature $T^f = 0.39$. The strictest inequality can be seen to be imposed in the limit $l \to \infty$.}\label{Fig:ldependence}
\end{figure}

Plots for a representative fast erasure of a Lorentzian excitation are shown in Fig. \ref{Fig:ldependence}. It is clear from the plots that the $l \to \infty$ limit gives the strictest QNEC inequality.

\subsection*{Fast erasure is impossible if $T^i = 0$}
\begin{figure}[t]
\includegraphics[width=0.4\textwidth]{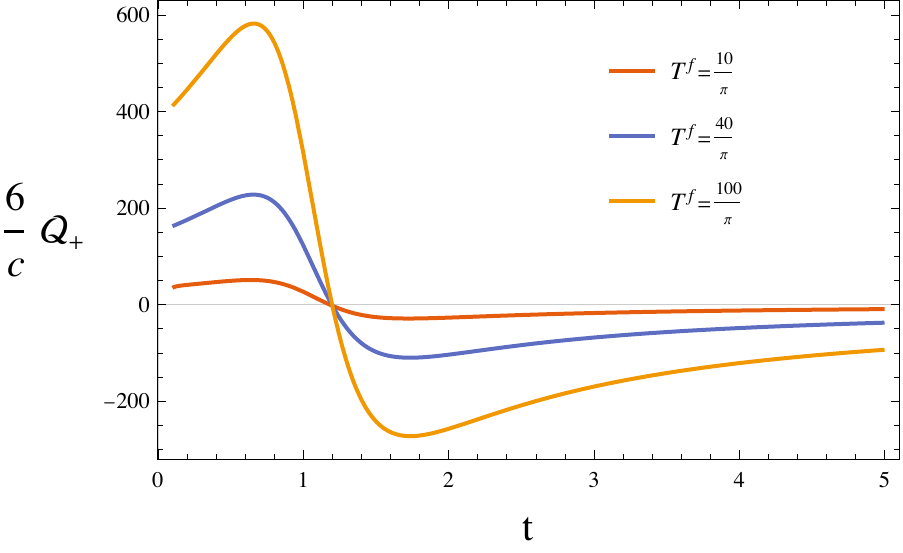}
\caption{$\mathcal{Q}_{+}$ is plotted for $T^i=0$, $a=1$, $b=1$ and various $T^f$. The violation becomes larger as we increase the final temperature. }\label{Fig:qnecd0}
\end{figure}

For $T^i = 0$ and arbitrary final temperature, initial amplitude and width we find that as $t \to \infty$
\begin{equation}
    \frac{6}{c} \mathcal{Q}_{+} = -4 \frac{\mu}{t}. 
\end{equation}
Thus QNEC is violated for any final temperature and therefore erasure is impossible if $T^i = 0$. This is illustrated in Fig. \ref{Fig:qnecd0}.

\subsection*{Method used to obtain the critical localization length $b_{c}$}
\begin{figure}[t]
\includegraphics[width=0.4\textwidth]{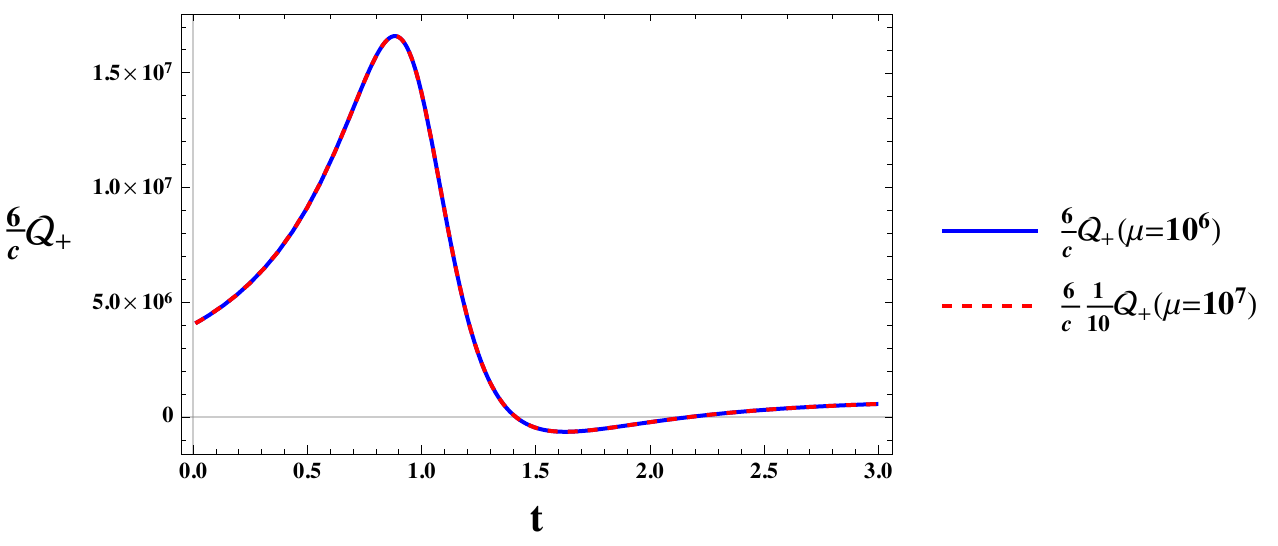}
\caption{A plot of the $\mathcal{Q}_{+}$ vs $t$ for $\frac{6}{c}s_{\rm info} = 2 \pi$, $b=0.3$, $T^i=\frac{0.2}{\pi}$, $x_{0}=-1$. We have chosen the entangling interval $x\geq x_0$ with $x_0 =-1$. For two choices of $\mu = 10^{6}$ and $\mu=10^{7}$ (note $\mu = \pi T^f$), we see that the $\mathcal{Q}_{+}(t)$ plot simply scales with $\mu$.}\label{Fig:bcalgo}
\end{figure}

As shown in figure 3 in the main text, for $b \geq b_{c}$, the fast erasure process is possible with a large enough $T^{f}$. For $b<b_{c}$ we find that QNEC is violated for a range of $s_{\rm info}$ and raising the final temperature simply scales the QNEC but doesn't cure the violation. This scaling behaviour is illustrated in Fig. \ref{Fig:bcalgo} and we use it to determine $b_c$ as described in the following algorithm.
\begin{enumerate}
    \item For a fixed $s_{\rm info}$ pick a $\mu$ (recall that the final state has temperature $T^f = \frac{\mu}{\pi}$).
    \item Vary the value of $b$ and find a $b_{c}$ such that QNEC is violated for all $b < b_{c}$.
    \item Increase $\mu$ to $ 10 \mu$.
    \item Repeat steps 2 and 3 to find a new $b_{c}$ that will be smaller than the older one.
    \item Repeat steps 2-4 until scaling behaviour is reached and $b_{c}$ cannot be decreased further by increasing $\mu$.
\end{enumerate}

\subsection*{Erasure of infinitely de-localized Mathieu excitations}
\begin{figure}[t]
\includegraphics[width=0.4\textwidth]{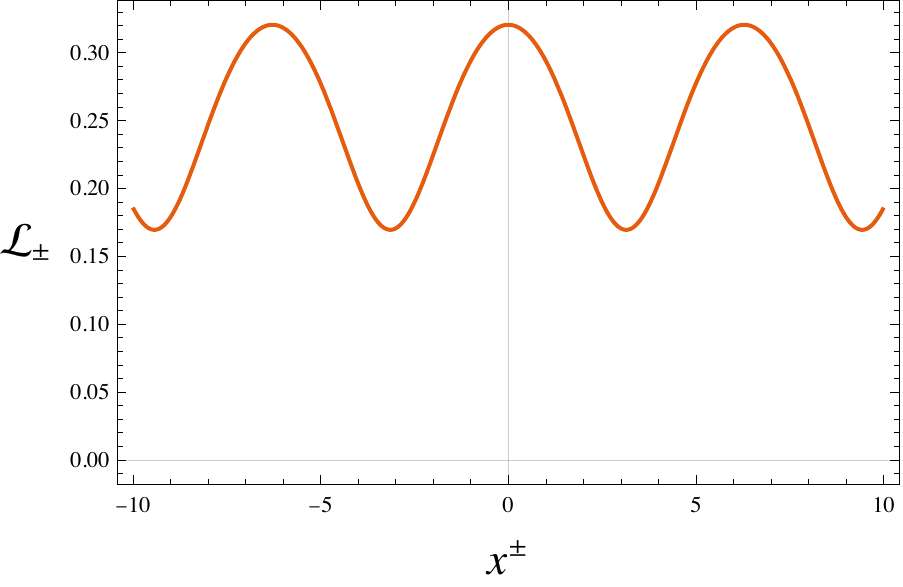}
\caption{A representative plot of $\mathcal L_{\pm}$ for $d=0.5$, $q=0.1$.}\label{Fig:mathieu}
\end{figure}

Using the cut and glue method described in the main text, here we study fast erasure of infinitely delocalized excitations to a thermal state with $T^f=\frac\mu\pi$. We consider excitations of the Mathieu form with
\begin{align}
    \mathcal L_{\pm}(x^{\pm}) &= d^{2} \alpha'(x^{\pm})-\frac{1}{2} \text{Sch}(\alpha(x^{\pm}),x^{\pm}),\\
    \alpha(x) &= x + q \sin(x), \qquad |q| \leq 1.
\end{align}
This particular form is chosen so that we can analytically obtain the uniformization map described in \cite{Kibe:2021qjy}. A representative plot of these excitations is shown in Fig. \ref{Fig:mathieu}. It is clear from the figure that these excitations are simply oscillations on top of a background temperature $T^i = \frac{d}{\pi}.$

The expression for $\mathcal{Q}_{+}$ at $t=0$ for a half line $x\geq x_{0}$ interval is
\begin{align}
    \mathcal{Q}_{+} &= \frac{c \mu ^3 q}{48 \left(\mu  q \cos \left(\mu  x_0\right)+1\right){}^2} \\
    &\times\big(-\mu  q \left(2 \sin \left(2 \mu  x_0\right)+3\right)+\mu  q \cos \left(2 \mu  x_0\right)\nonumber\\&\quad\quad -4 \sin \left(\mu  x_0\right)-2 \cos \left(\mu  x_0\right)\big)\nonumber
\end{align}
At $x_{0} = 0$ we get
\begin{equation}
    \mathcal{Q}_{+} = \frac{c}{6} \left(-\frac{\mu ^3 q}{4 \mu  q+4}\right).
\end{equation}
Therefore for $\mathcal{Q}_{+}\geq 0$ we must have 
\begin{equation}
    -\min\left(\frac{1}{\mu},1\right) \leq q \leq 0
    \label{eq:mathieucond1}
\end{equation}
At $x_{0} = \frac{3 \pi}{2 \mu}$ we get
\begin{equation}
    \mathcal{Q}_{+} = \frac{c}{6} \left( -\frac{1}{2} \mu ^3 q (\mu  q-1) \right).
\end{equation}
For $\mathcal{Q}_{+}\geq 0$ we must have in this case
\begin{equation}
    0 \leq q \leq \min\left(\frac{1}{\mu},1\right).
    \label{eq:mathieucond2}
\end{equation}
The two conditions \eqref{eq:mathieucond1} and \eqref{eq:mathieucond2} are inconsistent with each other and therefore QNEC cannot be satisfied for all $x_{0}$. An erasure of such an infinitely de-localized excitation is disallowed by QNEC.
\subsection*{Creation of Lorentzian excitations is disallowed}
\begin{figure}[t]
\includegraphics[width=0.4\textwidth]{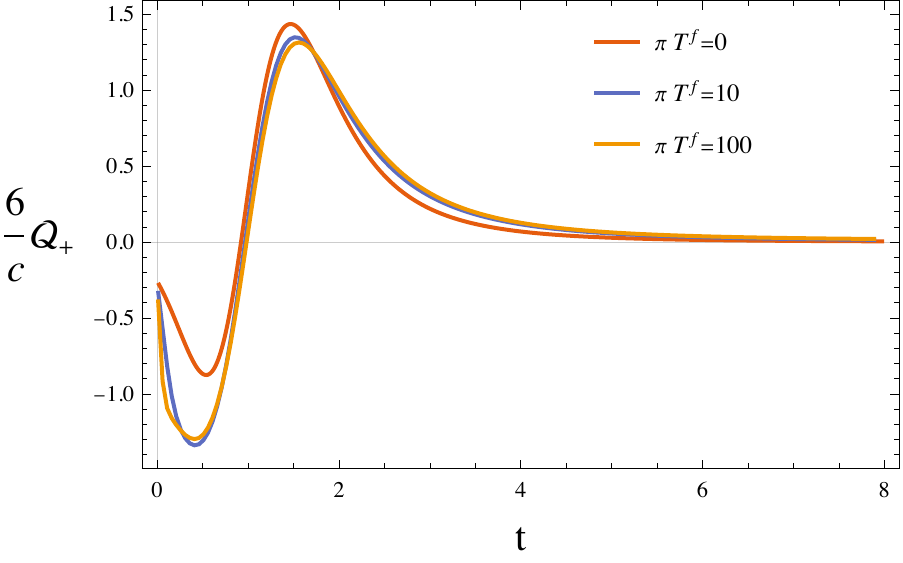}
\caption{$\mathcal{Q}_{+}$ for a quench transforming the vacuum state to a Lorentzian excitation with $x_{0} = -1$, $a=1$, $b=1$ for various $T^f$. The length of the entangling interval is taken to be $l=1000$, which is numerically the $l \to \infty$ limit.}\label{Fig:qnecoT0}
\end{figure}

In the main text we have analyzed the deletion of Lorentzian excitations. It is natural to ask if such excitations can be created from the vacuum or a thermal state via fast quenches of the form described in the main text. Fig. \ref{Fig:qnecoT0} shows a representative plot of $\mathcal{Q}_{+}$ for a quench that transforms the vacuum state to a Lorentzian excitation with various background temperatures. The QNEC in this case has to be computed numerically since the intersection points of the geodesic with the glueing hypersurfaces cannot be solved analytically. As we scale the length of the entangling interval we find that the plots for the QNEC with entangling lengths $l=1000$ and $l=5000$ coincide and we are therefore able to conclude that $l=1000$ is numerically the $l \to \infty$ limit. The plot in Fig. \ref{Fig:qnecoT0} shows that this transition is disallowed by QNEC and scaling the final temperature doesn't cure the QNEC violation but makes it worse. We scan the background temperature of the final state over the range $T^f = \frac{1}{\pi}$ to $T^f = \frac{10000}{\pi}$ and always find that QNEC is violated.

\begin{figure}[t]
\includegraphics[width=0.4\textwidth]{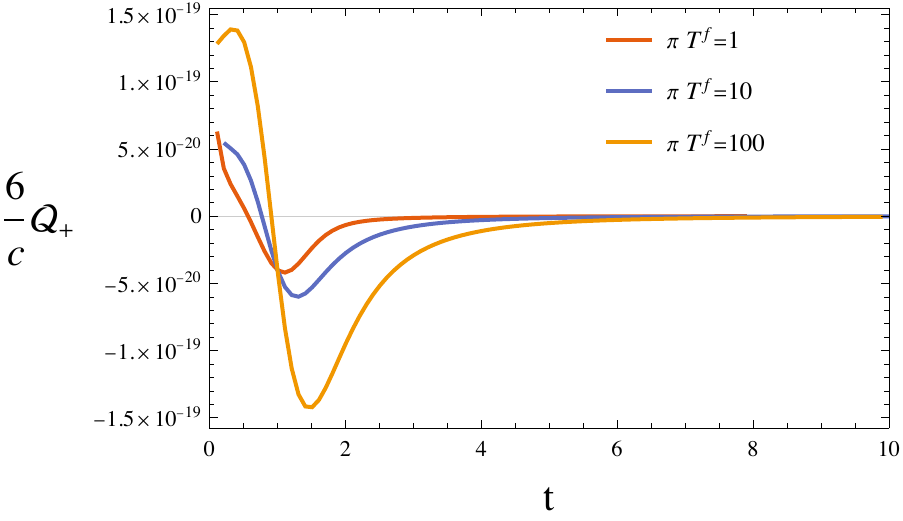}
\caption{$\mathcal{Q}_{+}$ for a quench transforming a thermal state with $T^i = \frac{1}{\pi}$ to a Lorentzian excitation with $x_{0} = -1$, $a=1$, $b=1$ and various $T^f$. The length of the entangling interval is taken to be $l=1000$, which is numerically the $l \to \infty$ limit. We work with numeric precision to 100 decimal places, therefore the $10^{-19}$ order of magnitude is not numerical noise.}\label{Fig:qnecbtzcreate}
\end{figure}

Similarly we can look at the creation of such excitations from an initial thermal state. Without loss of generality we choose the initial temperature $T^i = 1/\pi$. We fix the parameters of the excitation to be $a=1$, $b=1$ and scan over the background temperature. The numerical $l \to \infty$ limit is essentially $l=1000$ as described before. The plot for $\mathcal{Q}_{+}$ with $x_{0}=-1$ is shown in Fig. \ref{Fig:qnecbtzcreate} which shows that the QNEC violation doesn't improve but simply scales as we raise the background temperature.

Thus we find that the creation of such Lorentzian excitations of the CFT from the vacuum or a thermal state via an fast quench is disallowed by QNEC. It would be interesting to understand what leads to a violation of QNEC in these processes both geometrically and also from a CFT computation.

\bibliography{lumpserasure}

\end{document}